\documentclass[reprint,superscriptaddress,showpacs,showkeys,amsmath,amssymb,floatfix,aps,pra,longbibliography]{revtex4-1}
\usepackage{graphicx}

\usepackage{bm}

\usepackage[dvipsnames]{xcolor}
\usepackage{physics}
\usepackage{placeins}
\usepackage{graphics}
\usepackage{color}
\usepackage{hyperref}
\usepackage{multirow}
\usepackage{blindtext}
\usepackage{gensymb}
\usepackage{textcomp}
\begin{document}

\newcommand{\inc}[1]{{\color{blue} #1}}
\newcommand{\que}[1]{{\color{red} #1}}

\title{Radial Rashba spin-orbit fields in commensurate twisted transition-metal dichalcogenide bilayers}

\author{Thomas Naimer}\altaffiliation[These authors contributed equally.]{}
\email{thomas.naimer@physik.uni-regensburg.de}
\affiliation{Institute for Theoretical Physics, University of Regensburg, 93040 Regensburg, Germany}

\author{Paulo E. {Faria~Junior}}\altaffiliation[These authors contributed equally.]{}
\affiliation{Institute for Theoretical Physics, University of Regensburg, 93040 Regensburg, Germany}
\affiliation{Department of Physics, University of Central Florida, Orlando, Florida 32816, USA}
\affiliation{Department of Electrical and Computer Engineering, University of Central Florida, Orlando, Florida 32816, USA}

\author{Klaus Zollner}\altaffiliation[These authors contributed equally.]{}
\affiliation{Institute for Theoretical Physics, University of Regensburg, 93040 Regensburg, Germany}

\author{Jaroslav Fabian}
\affiliation{Institute for Theoretical Physics, University of Regensburg, 93040 Regensburg, Germany}

\begin{abstract}
In commensurate twisted homobilayers, purely radial Rashba spin-orbit fields can emerge. 
We employ first-principles calculations to investigate the band structures and the spin-orbit fields close to the high-symmetry points $K$ and $\Gamma$ of several commensurate twisted transition-metal dichalcogenide homobilayers: WSe$_2$, NbSe$_2$, and WTe$_2$.
The observed in-plane spin textures are mostly radial, and the main features are successfully reproduced using a model Hamiltonian based on two effective mass models including spin-orbit coupling, and a general (spin-conserving) interlayer coupling.
Extracting the model Hamiltonian parameters through fitting of several twisted supercells, we find a twist angle dependency of the magnitude of the radial Rashba field, which is symmetric not only around the untwisted cases ($\Theta=0^\circ$ and $\Theta=60^\circ$), but also around $\Theta=30^\circ$. Furthermore, we observe that the interlayer coupling between the $K/K'$-points of the two layers decreases with the increase of the size of the commensurate
supercells. Hence, peaks of high interlayer coupling can occur only for twist angles, where small commensurate supercells are possible. Exploring different lateral displacements between the layers, we confirm that the relevant symmetry protecting the radial Rashba is 
an in-plane 180$^\circ$ rotation axis. We additionally investigate the effects of atomic relaxation and modulation of the interlayer distance. {Our calculations on WTe$_2$ bilayers  show that their lack of $C_3$ symmetry results in spin textures that are neither radial nor tangential.}
Our results offer fundamental microscopic insights that are particularly relevant to engineering spin-charge conversion schemes based on twisted layered materials.
\end{abstract}

\pacs{}
\keywords{transition-metal dichalcogenides, twistronics, spin-orbit coupling}
\maketitle

\section{Introduction}
Transition-metal dichalcogenides (TMDCs) are a class of layered van der Waals materials that have a variety of applications including valleytronics~\cite{Xiao2012PRL, Liu2019:NanRes:TMDCValleytr,Schaibley2016:NRW:valleytronicsReview,Langer2018:Nat:valleytronics}, straintronics~\cite{Blundo2020:PRR:TMDCstrainbubbles,Zollner2019:strain,FariaJunior2022NJP}, optoelectronics~\cite{Thakar2020:TMDCoptoelectronicsReview,Mak2016:NP:TMDCoptoelectronics} and  spintronics~\cite{Zutic2004:RMP,Avsar2020:GrapheneSpintronicsReview}. Especially their 2D forms as mono- or bilayers exhibit interesting, versatile physics. They often come in a 2H configuration with a hexagonal unit cell, featuring parabolic bands both at the $\Gamma$- and the $K$-point. For the monolayer, the bands at $K$ are usually split by a strong spin-orbit coupling (SOC) of the valley-Zeeman type with strong out-of-plane polarized spins. Nevertheless, when the horizontal mirror symmetry is broken, Rashba SOC\cite{Bychkow1984:JETPL:Rashba} can also arise in these bands, introducing an in-plane spin texture. This breaking of symmetry can occur both externally (through an interface or an external electric field) or internally (through a lateral shift between the layers or a twisting of the layers).
The in-plane spin texture induced by the Rashba SOC can be used for spin manipulation and relaxation in spintronics\cite{Zutic2004:RMP, Avsar2020:GrapheneSpintronicsReview}. One major application, utilizing the Rashba SOC, is charge-to-spin conversion through the Rashba Edelstein effect\cite{Edelstein1990:SSC}.

The breaking of vertical mirror symmetry that naturally occurs in twisted multilayer systems can introduce a radial component to the (usually purely tangential) in-plane Rashba spin texture. Such 'radial Rashba' spin-orbit fields are interesting to the field of spintronics as they can enable unconventional charge-to-spin conversion\cite{Ingla_Aynes2022:expradialrashba2,Camosi_2022:2DM:expradialrashba3,Ontonso2023:PRA:expradialrashba4,Peterfalvi2022:radialrashba,Yang2024:NM:GRTMDCradialexp}. Furthermore, it can be used in Josephson junctions in combination with an external magnetic field to enable the radial superconducting diode effect\cite{Costa2023:PRB:superconddiode,Baumgartner2022:JP:superconddiode,Baumgartner2022:NN:superconddiode,Kang2024:PRB:superconddiodeDenis}. The first density functional theory (DFT) calculations observing such radial Rashba were performed on graphene-based heterostructures\cite{Naimer2021:paper1,Naimer2023:paper2,Naimer2024:paper3,Zollner2023:PRB,Lee22:PRB:radialrashba3} and usually find only rather small deviations from the tangential pattern.
Further investigations of commensurate twisted graphene homobilayers and proximitized graphene structures\cite{Frank2024:PRB:pureradialRashba}, however, revealed a purely radial Rashba spin-orbit field{\cite{Frank2024:PRB:pureradialRashba,Peterfalvi2022:radialrashba,Gatti2020:PRL:chiraltellurium,Sakano2020:PRL:chiraltellurium2,Lin2022:PRB,Krieger2024:NC,GosalbezMartinez2023:3Dsymmetriesradial2,deSousa2025:radialRashba}}, which arises due to the coupling between the 'hidden Rashba' spin-orbit fields of the two layers.

In this paper, we systematically explore the emergence of the radial Rashba SOC in twisted TMDC homobilayers.
We perform first principles calculations for several 
homobilayer supercells, exploring different materials (WSe$_2$, NbSe$_2$ and WTe$_2$), shifting configurations, twist angles, and interlayer distances. Since the commonly used models of twisted TMDC homobilayers~\cite{Wu2017:PRL:twTMDC5} are designed for small twist angles, here we employ an effective model Hamiltonian, similar to Ref.~\cite{Frank2024:PRB:pureradialRashba}, but replacing graphene with (an effective mass model of) TMDC. This model is designed to work for generic twist angles, as long as the underlying supercell is commensurate, which is naturally satisfied in first-principles calculations due to periodic boundary conditions.
We extract the spin-orbit fields around both $K$ and $\Gamma$ and draw conclusions about the twist angle dependencies of the parameters. Additionally, we establish that an in-plane 180$^\circ$ rotation symmetry is crucial for the emergence of the purely radial Rashba.

The paper is structured in the following way. Secs.~\ref{Sec:globalBS}-\ref{Sec:SOF} discuss hexagonal homobilayer structures  (WSe$_2$ and NbSe$_2$). In Sec.~\ref{Sec:globalBS} we explore the different possible commensurate DFT supercells and the three different types of band backfoldings that can occur. Sec.~\ref{Sec:modelHam} introduces the model Hamiltonian we use for describing these cases. A more detailed look at the spin-orbit fields derived from DFT as well as the parameters extracted from the model Hamiltonian fits can be found in Sec.~\ref{Sec:SOF}. Finally, Sec.~\ref{Sec:WTe} is dedicated to a discussion of WTe$_2$, which --- due to its rectangular supercell and lack of C$_3$ symmetry --- cannot be properly described by our model Hamiltonian. In App.~\ref{App:relationtoconttw} we discuss how the interlayer coupling can be described as continuous function of the twist angle.  App.~\ref{App:interlayerdist} presents a study on  the effects of varying interlayer distance. Computational details of the DFT calculations and fitting are giving in App.~\ref{App:compdet}.

\section{Global band structures and backfolding}
\label{Sec:globalBS}
\begin{figure}
    \centering
    \includegraphics[width=0.99\linewidth]{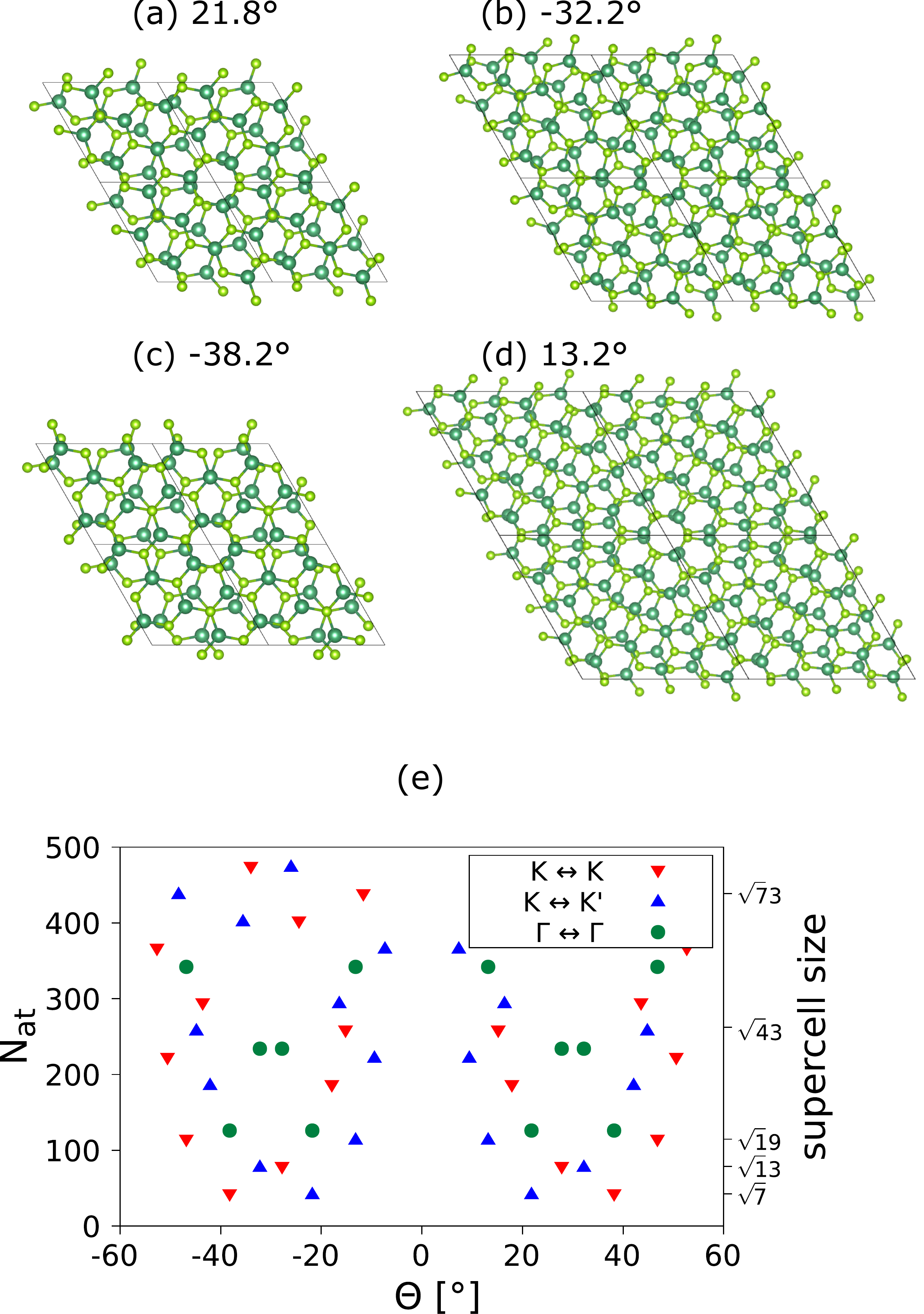}
    \caption{(a)-(d) Top view of exemplary investigated commensurate supercells. In (e) possible twisted commensurate bilayer supercell sizes $N_{at}$ (number of atoms) are indicated for different twist angles $\Theta$. For each supercell data point, we color-code where the $K/K'$ points of the layers fold back to: For red downward-pointing triangles, the $K$ points of both layers fold back to the same point in the supercell's FBZ ($K\leftrightarrow K$). For blue upward-pointing triangles, $K$ of layer 1 and $K'$ of layer 2 fold on top of each other and vice versa ($K\leftrightarrow K'$). For green dots all $K$- and $K'$-points of both layers fold to $\Gamma$. The last cases are not discussed in this paper and only listed for completeness.}
    \label{Fig:strucsbackfolding}
\end{figure}

In order to calculate the properties of twisted structures using DFT, we need to construct commensurate supercells that satisfy in-plane periodic boundary conditions. For homobilayers of materials with hexagonal unit cells (WSe$_2$ and NbSe$_2$), there are twist angles where this can easily be done without the introduction of strain for moderate supercell sizes \cite{Uchida2014:PRB}. A comprehensive scatter plot showing feasible supercells and some specific structures used throughout the paper are shown in Fig.~\ref{Fig:strucsbackfolding}. For each supercell there is a partner supercell, which can be obtained by twisting one of the layers by 60$^\circ$. The two partner angles with the smallest corresponding supercells are 21.8$^\circ$ and -38.2$^\circ$. Supercells with negative twist angles can easily be related to their counterparts (e.g. -38.2$^\circ$ to 38.2$^\circ$). They exhibit Rashba fields with same magnitude, but opposite sign, as they are related by an in-plane mirror symmetry ($z\rightarrow -z$). Additionally their $K$- and $K'$-points are swapped.
If not mentioned otherwise, we twist the layers around a common twisting axis going through the 'Hollow' position (i.e. in the middle of the hexagon formed by the metal and chalcogen atoms) of both layers. We define the supercell size of a certain supercell as $s=a_s/a$, where $a_s$ is the lattice constant of the supercell and $a$ is the lattice constant of the primitive monolayer unit cell.

\begin{figure}
    \centering
    \includegraphics[width=0.89\linewidth]{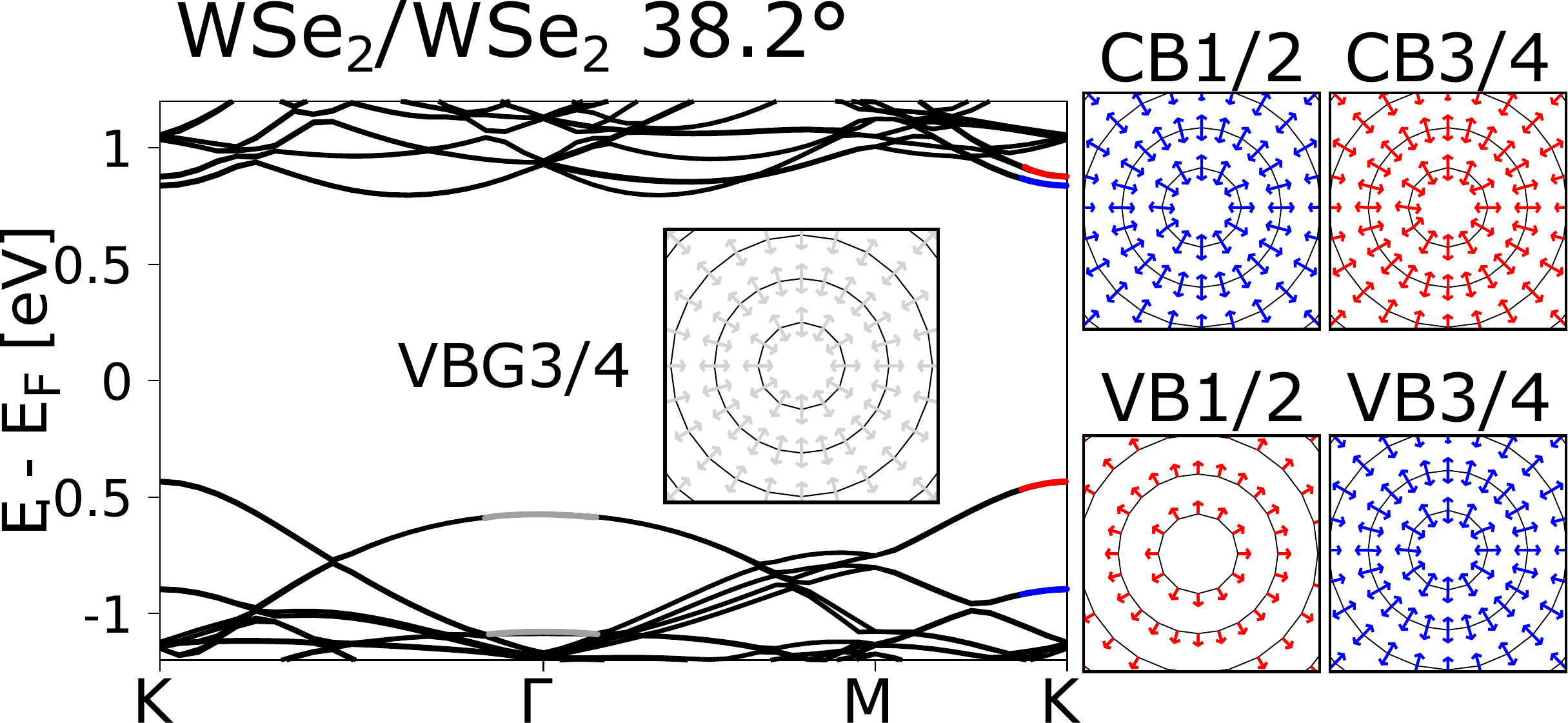}
    \caption{Band structure along high symmetry points and in-plane spin textures around $K$ and $\Gamma$ of the 38.2$^\circ$ twisted WSe$_2$ homobilayer, which shows a $K\leftrightarrow K$ backfolding. Out-of-plane spin is color coded from blue (spin down) over grey to red (spin up). }
    \label{Fig:globalBSWSe}
\end{figure}

\begin{figure}
    \centering
    \includegraphics[width=0.99\linewidth]{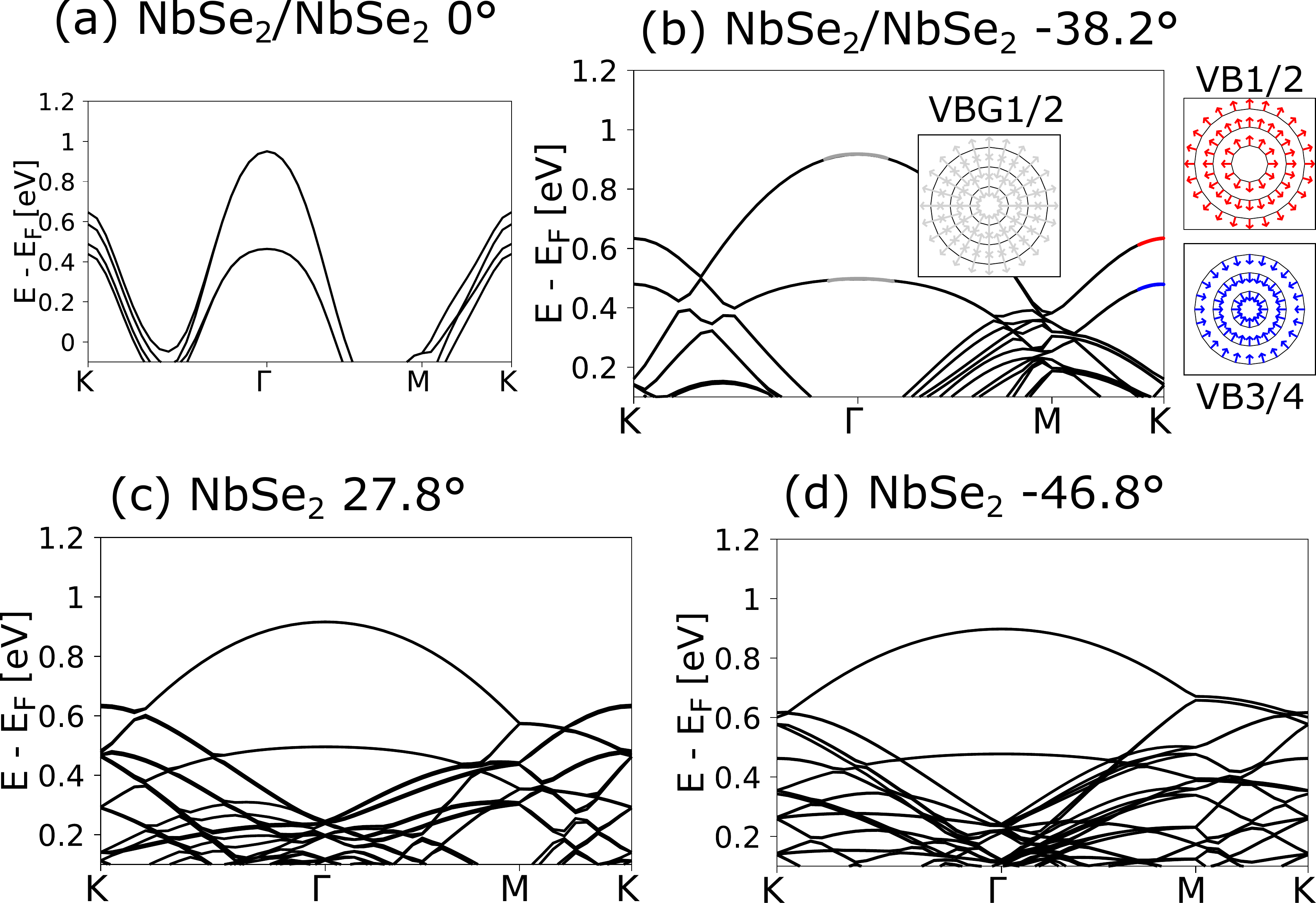}
    \caption{Band structures along high symmetry points of exemplary NbSe$_2$ homobilayers, all of which exhibit a $K\leftrightarrow K$ backfolding. (a) shows the untwisted case with a lateral shifting position, in which a metal atom of one layer resides on top of a chalcogen atom of the other layer. The split bands are layer polarized due to the breaking of the in-plane mirror symmetry.
    In (b), we additionally show the in-plane spin textures around $K$ and $\Gamma$ for the -38.2$^\circ$ case. (c) and (d) show the band structures of 27.8$^\circ$ and -46.8$^\circ$ cases. Out-of-plane spin is color coded from blue (spin down) over grey to red (spin up). }
    \label{Fig:globalBSNbSe}
\end{figure}

The band structures and spin textures are calculated using DFT with the computational details given in App.~\ref{App:compdet}.
In Fig.~\ref{Fig:globalBSWSe} and Fig.~\ref{Fig:globalBSNbSe}~(b), we present the band structures along high-symmetry lines and spin textures around $K$ and $\Gamma$ of the $\pm38.2^\circ$ twisted WSe$_2$/WSe$_2$ and NbSe$_2$/NbSe$_2$ band structures, respectively. 
{For the case of NbSe$_2$, we additionally show the band structures of the untwisted structure in Fig.~\ref{Fig:globalBSNbSe}~(a) and  further twisted structures in Fig.~\ref{Fig:globalBSNbSe}~(c)-(d).}
At the $\Gamma$-point, the bands of the two layers are strongly hybridized as also observed e.g. in Ref.~\cite{rosa2025:APR} for MoSe$_2$ homobilayer systems. Naturally, there is no valley-Zeeman splitting to be observed at $\Gamma$. The bands show no out-of-plane spins, the splitting between them is rather caused by the interaction between the layers. This is illustrated by the fact that the splitting scales with the interlayer distance.

The key to understanding the bands at the $K$-point of the supercell's first Brillouin zone (FBZ) is the backfolding of the bands of the two layers. For the commensurate structures we investigate, there are three options\cite{Naimer2024:thesis}, which we color code for all possible commensurable supercells in Fig.~\ref{Fig:strucsbackfolding} d):
\begin{enumerate}
    \item $K\leftrightarrow K$ (red downward triangles): The $K$-point of layer 1 folds on top of the $K$-point of layer 2 (and the same for $K'$). In this case the spin-up (and spin-down) bands of both layers are always at the same energy. Hence, their out-of-plane spin-polarization is kept intact.
    \item $K\leftrightarrow K'$ (blue upward triangles): The $K$-point of layer 1 folds on top of the $K'$-point of layer 2 and vice versa. In this case, the spin-up band of layer 1 and the spin-down band of layer 2 can easily interact, which can lead to a suppression of the out-of-plane spins.
    \item $\Gamma\leftrightarrow \Gamma$ (green dots): The $K$- and $K'$-points of both layers folds to the $\Gamma$-point. Although this option is technically possible, it is always a 'supercell of a supercell', i.e. there is always a smaller supercell (which can be categorized in one of the former categories) at the same twist angle, representing the same physics. In this case the bands would fold on top of each other, but only those already connected in the smaller supercell could interact. 
\end{enumerate}
In this paper we are only considering supercells of the first two cases. As the used supercells are furthermore the smallest possible supercells at the specific angle, we can be sure that the backfolded bands are also coupled to each other via generalized Umklapp processes\cite{Koshino2015:TwistTBBasic,Naimer2024:thesis}. For the bands stemming from the $\Gamma$-point, we do not need such an argumentation as they are always directly coupled. 
Two partner angles will always exhibit opposing backfolding cases, e.g. 21.8$^\circ$ has $K\leftrightarrow K'$, while -38.2$^\circ$ has $K\leftrightarrow K$ backfolding. 

\section{Model Hamiltonian}
\label{Sec:modelHam}
In all commensurate supercells we employ, we find nearly parabolic bands (from $K$ or $\Gamma$) of the individual layers' FBZs folding back on top of each other and interacting with each other.
In order to describe these interacting parabolic bands we employ a model Hamiltonian.
It consists of two effective mass models for the two layers (including SOC terms), which are interacting by a general (spin-conserving) interlayer coupling. This might be viewed as a SOC-including version of the moiré  Hamiltonians used in Refs.~\cite{Devakul2021:twTMDC3,Wu2019:PRL:twTMDC4,Wu2017:PRL:twTMDC5,Wu2018:PRB:twTMDC6}, just without the Moire potential. However, these models additionally describe a shift in $k$-space between the layers' bands, which comes from a small twist between the layers. Our model Hamiltonian, on the other hand, is aimed to describe the case where the two layers' bands are lying directly on top of each other, as it happens in all commensurate supercells. This is the same as the Hamiltonian in Ref.~\cite{Frank2024:PRB:pureradialRashba} does for graphene Dirac cones. Although, in principle, our model can also describe cases of commensurate supercells with small twist angles, in these cases the supercells will be large and hence the interaction between  the directly overlapping bands will be small (see App.~\ref{App:relationtoconttw}). Rather, the physics of these small angle cases will be dominated by the interaction between the slightly mismatched $K$-points (mini Brillouin zone).

The model Hamiltonian consist of an orbital part and two SOC terms:
\begin{align}
\label{Eq:Ham}
    H(\mathbf{k})=H_{orb}(\mathbf{k})\otimes s_0+H_{VZ}+H_{R}(\mathbf{k}).
\end{align}
Here, {$\otimes$ is the Kronecker product and $s_0$ is the zeroth Pauli matrix (identity matrix) for the spin degree of freedom, as the orbital part does not include SOC}; $\mathbf{k}$ is measured either from $K$ or from $\Gamma$, depending on which bands are to be described.
 The orbital part of the Hamiltonian 
\begin{align}
H_{orb}(\mathbf{k})=
    \begin{pmatrix}
    \frac{\hbar^2(\mathbf{k})^2}{2m_{\text{eff}}}&  w\\
    w  &  \frac{\hbar^2(\mathbf{k})^2}{2m_{\text{eff}}}
\end{pmatrix}
\end{align}
includes the parabolic bands (with effective mass $m_{\text{eff}}<$0 for valence or $m_{\text{eff}}>$0 for conduction bands) of the two layers in the diagonals.
The interlayer coupling is parametrized by $w$ taken to be real (we checked that adding a phase does not change any relevant physics).

When adding SOC, two SOC terms need to be considered: the Rashba and valley-Zeeman type SOC. 
The latter one is given as
\begin{align}
\label{Eq:VZ1}
H_{VZ}=H_{VZ,K\leftrightarrow K}=\lambda_{VZ}\sigma_0\otimes s_z=\lambda_{VZ}\otimes
    \begin{pmatrix}
    s_z&  0\\
    0&  s_z
\end{pmatrix}
\end{align}
or
\begin{align}
\label{Eq:VZ2}
H_{VZ}=H_{VZ,K\leftrightarrow K'}=\lambda_{VZ}\sigma_z\otimes s_z=\lambda_{VZ}\otimes
    \begin{pmatrix}
    s_z&  0\\
    0&  -s_z
\end{pmatrix},
\end{align}
depending on the backfolding case. Here, the Pauli matrices $s_i$ describe the spin degree of freedom, while $\sigma_i$ describe the layer degree of freedom. Eq.~\ref{Eq:VZ1} corresponds to the case $K\leftrightarrow K$, where the valley-Zeeman splittings of the two layers have the same sign.  Eq.~\ref{Eq:VZ2} corresponds to the case where $K\leftrightarrow K'$, where the valley-Zeeman splittings of the two layers have the opposite sign.

For a single layer, the Rashba SOC can be described with the typical semiconductor Rashba Hamiltonian\cite{Kormanyos2014:PRX}. In order to describe the effect of the hidden Rashba of the twisted layers we need to implement a twisted spin texture. To this end the spin texture of each layer is twisted by $\Phi$ in opposite directions, where $\Phi$ is the Rashba angle, similar to the Rashba angle of graphene/TMDC heterostructures~\cite{Naimer2021:paper1,Peterfalvi2022:radialrashba,Veneri22:PRB:radialrashba2}:
\begin{align}
    H_{\text{R,mono}}(\mathbf{k},\Phi)=\exp(\frac{i\Phi s_z}{2})\Big[\lambda_R(s_yk_x-s_xk_y)\Big]\exp(-\frac{i\Phi s_z}{2}).
\end{align}
A Rashba angle of zero ($\Phi=0$) corresponds to the case of conventional Rashba (with spins tangential to the momentum), while $\Phi=\pm90^\circ$ corresponds to an unconventional Rashba with purely radial in-plane spins.
The total Rashba SOC of the system is then described by two monolayer Hamiltonians with opposite signs and opposite $\Phi$:
\begin{align}
H_{R}(\mathbf{k},\Phi)=
    \begin{pmatrix}
    H_{\text{R,mono}} (\mathbf{k},\Phi) & 0 \\
    0 & -H_{\text{R,mono}} (\mathbf{k},-\Phi)
\end{pmatrix}.
\end{align}

Note that in Ref.~\cite{Frank2024:PRB:pureradialRashba} it is assumed that $\Phi$ can be expressed via the twist angle as $\Phi=\Theta/2$. Such an assumption is reasonable if the focus is on the qualitative emergence of a radial spin texture and not on a quantitative evaluation of the magnitude of the radial Rashba. There are symmetry rules for the mapping between the twist angle $\Theta$ and the Rashba angle $\Phi$, which are similar to the ones valid for graphene/TMDC heterostructures. 
\begin{align}
\label{Eq:sym}
    \Phi(\Theta+120^\circ)&=\Phi(\Theta)\\
    \label{Eq:sym2}
    \Phi(-\Theta)&=-\Phi(\Theta)
\end{align}
As a consequence $\Phi(\Theta=0^\circ)=\Phi(\Theta=60^\circ)=0+n\pi$ for $n\in\mathbb{Z}$.

\begin{figure}
    \centering
    \includegraphics[width=0.99\linewidth]{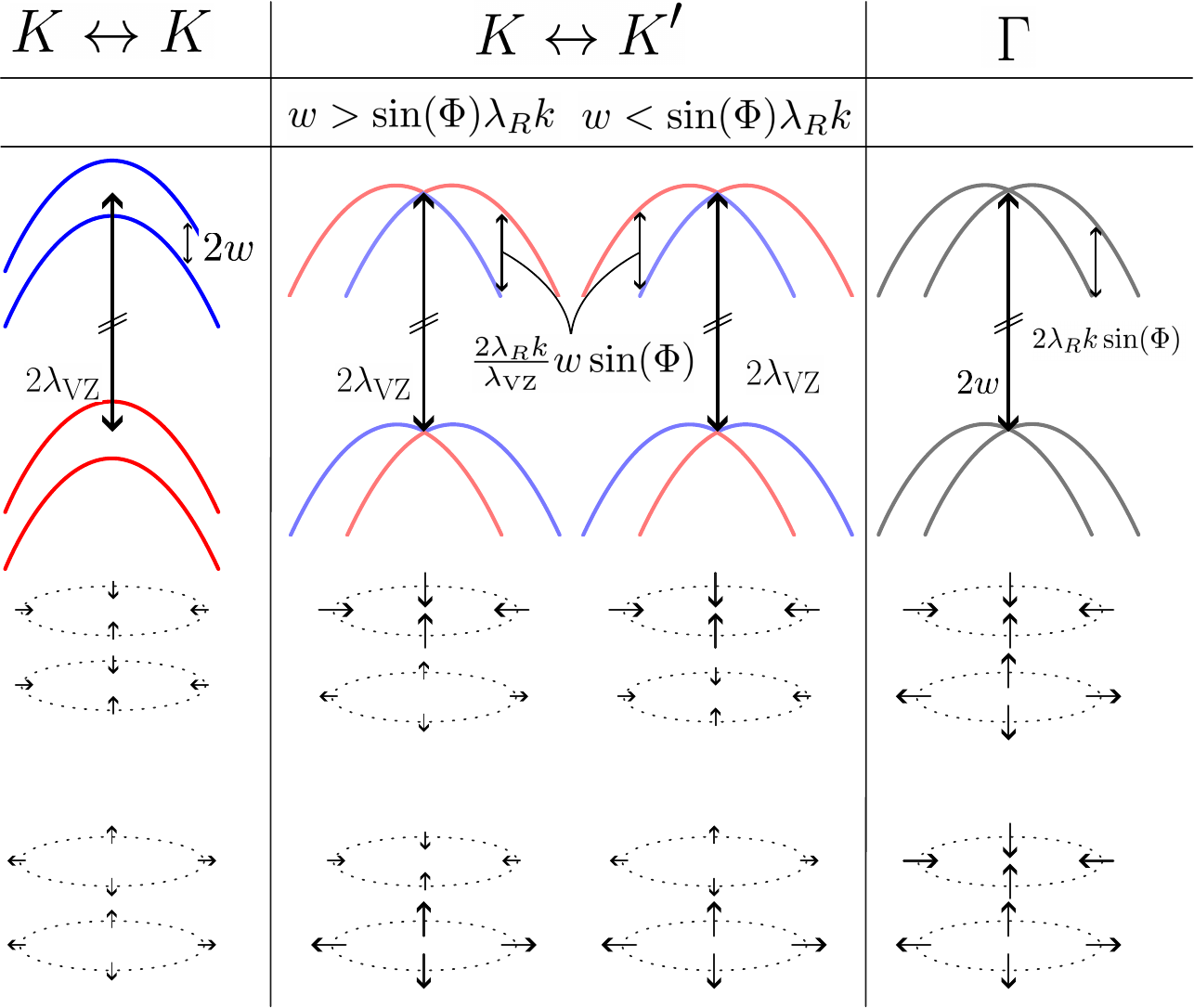}
    \caption{Schematic depiction of the bands of the twisted model Hamiltonian for different cases: backfolding of both $K$-points of the layers on top of each other ($K\leftrightarrow K$), backfolding of $K$-point of layer 1 on top of $K'$ of layer 2 ($K\leftrightarrow K'$) and when describing the $\Gamma$ bands. In the latter case the backfolding is irrelevant. We depict for all cases the parabolic (valence) bands and their color coded spin-$z$ expectation values with analytical expressions of the splittings (with assumption $\lambda_{VZ}>>w$ for the first two cases and $\lambda_{VZ}<<w$ for the third case). Finally, we also depict the in-plane spin textures of the four bands ordered by energy.}
    \label{Fig:model}
\end{figure}

Let us now briefly discuss which spin textures this model Hamiltonian displays for different ranges of parameters. Within the model, a system with twisted ($\Theta\neq0,\Phi\neq0)$, interacting ($w\neq0$) layers, which can provide Rashba SOC ($\lambda_R\neq0$), the resulting spin texture is always radial, as the tangential parts of the spin textures of the two layers cancel out. This is the same principle as for twisted bilayer graphene structures~\cite{Frank2024:PRB:pureradialRashba}. The magnitude and sign of this emerging radial in-plane spin texture, the splittings of the bands, as well as the spin-$z$ and layer hybridization, are, however, dependent on the specific case and parameter range. The typical cases we find are:
\begin{enumerate}
    \item $K\leftrightarrow K$ backfolding with dominant valley-Zeeman SOC $\lambda_{VZ}>>\lambda_R$: states of the two layers with same spin-$z$ fold on top of each other, keeping the strong spin-$z$ polarization (caused by the strong $\lambda_{VZ}$) and the layer polarization intact. The two adjacent bands with same spin form a band pair, which are split by a uniform splitting of $2w$. Within this band pair, the sign of the radial Rashba (whether it is pointing inward or outward) is always uniform. The magnitude of the radial in-plane structure (i.e the in-plane spin expectation value $\sqrt{\langle s_x\rangle^2+\langle s_y\rangle^2}$) scales with $\frac{\lambda_Rk}{2\lambda_{VZ}}\sin(\Phi)$.
    
    \item $K\leftrightarrow K'$ backfolding with dominant valley-Zeeman SOC $\lambda_{VZ}>>\lambda_R$: states of the two layers with opposite spin-$z$ fold on top of each other. Hence,  the spin-$z$ and layer polarization are mostly lifted. The splitting within the  band pairs is $2\cdot\frac{\lambda_Rk}{\lambda_{VZ}}\cdot w\cdot \sin(\Phi)$. 
    Regarding the sign of the radial Rashba, two subregimes arise. For small interlayer couplings ($w<<\lambda_Rk$) or large $k$ radii, the two adjacent bands are aligned. With rising $w$ one of the bands of each band pair exhibits linearly decreasing magnitude of the radial spin texture until it switches sign at $w=\sin(\Phi)\lambda_Rk$. After this point ($w>>\lambda_Rk)$, the two adjacent bands exhibit opposite signs of their radial Rashba. The magnitude of the radial in-plane structure scales with $\frac{w}{2\lambda_{VZ}}\pm\frac{\lambda_R k\sin(\Phi)}{2\lambda_{VZ}}$.

    \item Dominant interlayer coupling $w>\lambda_Rk$ and absent valley-Zeeman SOC $\lambda_{VZ}$ (describing the bands at $\Gamma$): Here, the two band pairs are separated by the interlayer coupling $w$ rather than the valley-Zeeman SOC. The splitting within each band pair is $2\lambda_Rk\sin(\Phi)$. The radial Rashba of the two adjacent bands of one band pair always show opposite signs.
\end{enumerate}
We illustrate the three cases in Fig.~\ref{Fig:model}.

\section{Spin-orbit fields from first principles}
\label{Sec:SOF}
\begin{figure*}
    \centering
    \includegraphics[width=0.99\linewidth]{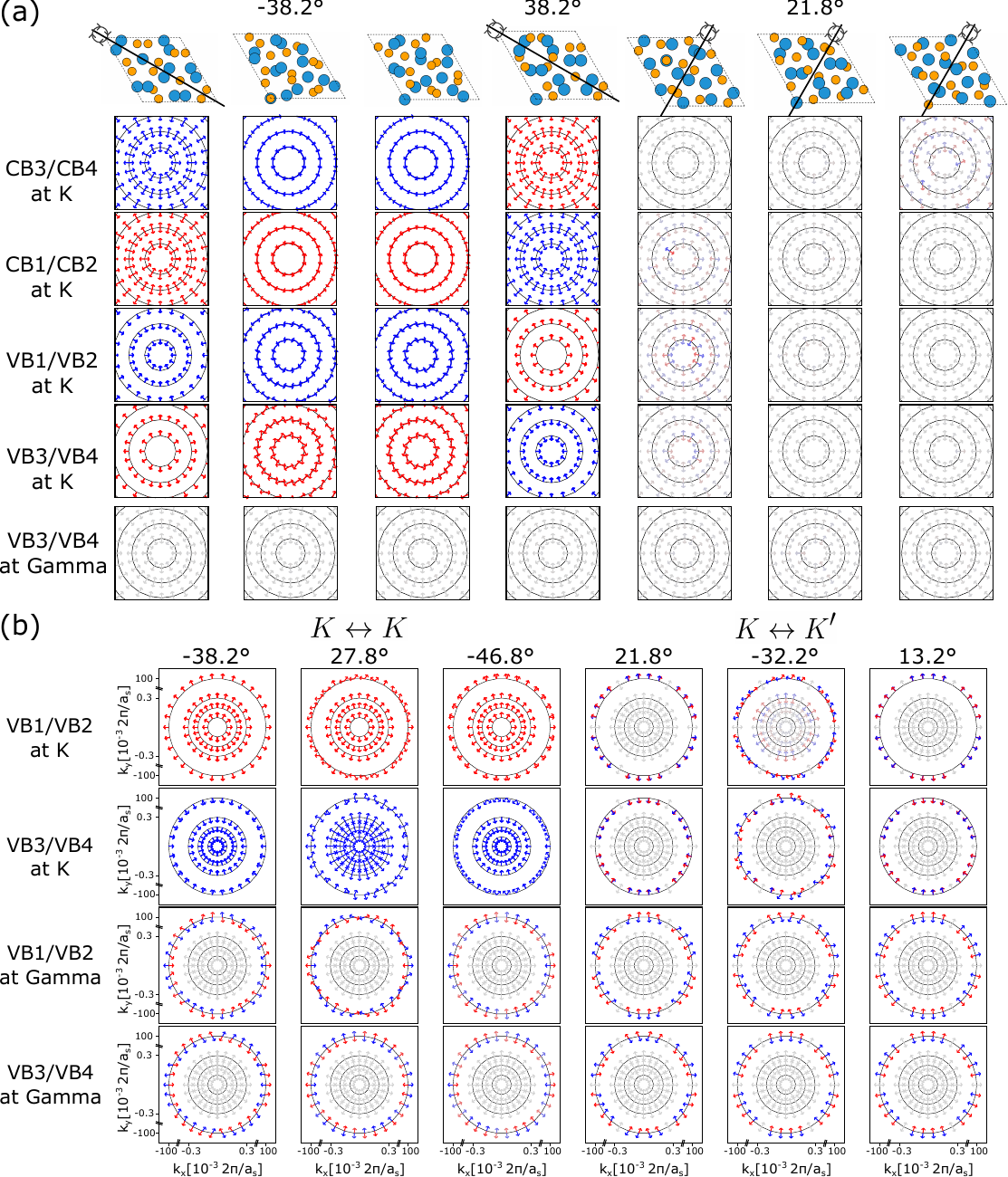}
    \caption{DFT extracted spin-orbit fields around $K$ and $\Gamma$ for (a) twisted WSe$_2$ and (b) twisted NbSe$_2$ homobilayers. In (a) the maximal range of the rings is at a $k$-radius of $0.3^\cdot10^{-3}\frac{2\pi}{a_s}$, where $a_s$ is the supercell lattice constant. In (b) one additional ring (with a $k$ radius of $0.1\frac{2\pi}{a_s}$) extends this range significantly. {At each $k$ point, two arrows are plotted for two adjacent bands, e.g. VB1 and VB2.}  Direction (not magnitude) of the in-plane spin textures are represented by the arrows. The out-of-plane spins are color coded from red over grey to blue.  In (a) we additionally show a top view of the supercell (with the 180$^\circ$ rotation axis, if present) for each different lateral shifting. Twist angles, backfolding scenarios and lateral shifts are defining the columns. The rows are defined by the band pairs (sorted by energy) and the relevant close $k$-point ($K$ or $\Gamma$). VB1/VB2 are omitted in (a) due to problems with determining spin expectation values unambiguously, due to a near-degeneracy of the bands.}
    \label{Fig:SOF}
\end{figure*}

Let us now examine the spin-orbit fields close to the relevant high-symmetry points. We show these spin-orbit fields for different supercells ($\Theta=21.8^\circ$, $\Theta=-38.2^\circ$, $\Theta=-32.2^\circ$, $\Theta=27.8^\circ$, $\Theta=-46.8^\circ$ and $\Theta=13.2^\circ$), different lateral shifts, different materials (WSe$_2$ and NbSe$_2$), different bands (valence and conduction bands) and around different high-symmetry points ($K$ and $\Gamma$) in Fig.~\ref{Fig:SOF}. In most of the cases the observed spin textures match well with the predictions by the model Hamiltonian. Especially the alignment of the spins when moving from small $k$ radii ($w>>\lambda_Rk$) to large ones ($w<<\lambda_Rk$) in the $K\leftrightarrow K'$ case can be observed in the DFT data (see Fig.~\ref{Fig:SOF} (b)). {It is apparent that, when going to larger k-radii, the spin texture is no longer perfectly isotropic, as assumed in the model. Rather, the spin texture here is often defined by the nodal lines arising from the combination of the $180^\circ$ rotational symmetry and the $C_3$ symmetry. However, also for small k-radii there are cases} where the model Hamiltonian fails to describe the spin textures properly. This for example is the case for the conduction bands of $\pm38.2^\circ$ twisted WSe$_2$ bilayers. Here, the smaller value of $\lambda_{VZ}$ in combination with a perturbation from nearby bands might be responsible for the deviation from the model.  {Generally, we also observe in DFT the splittings within the bands pairs predicted by the model Hamiltonian (especially their dependence on the backfolding scenario, see Fig.~\ref{Fig:model}). Only the magnitude of the splittings for the $K\leftrightarrow K'$ case is larger in DFT than the one predicted by the model (see Fig.~\ref{Fig:fitting} in App.~\ref{App:compdet}).}

For WSe$_2$, we performed calculations on multiple shifting configurations, while for NbSe$_2$ we utilized supercells with multiple different twist angles. Hence, we split the discussion into two parts: 
\begin{enumerate}
    \item Discussion of the shifting degree of freedom and relevant symmetries using WSe$_2$ (Fig.~\ref{Fig:SOF} (a))
    \item Discussion of the twist angle dependency of the extracted parameters using NbSe$_2$ (Fig.~\ref{Fig:SOF} (b) and Fig.~\ref{Fig:extractedparams})
\end{enumerate}
Nevertheless, the arguments are general and not restricted to the particular TMDC.

\subsection{Relevance of the in-plane 180$^\circ$ rotational symmetry}
One degree of freedom that has so far not been discussed is the lateral shifting between the two layers. This is equivalent to considering different axes around which the twisting occurs.  Examining twisted supercells with different lateral shiftings (Fig.~\ref{Fig:SOF} (a)), we consistently find that the purely radial Rashba spin textures can only be maintained in systems with an intact in-plane 180$^\circ$-rotation axis.
This can be understood in the following way:  The 180$^\circ$-rotation operation can be considered as a combination of a vertical mirror operation and a horizontal mirror operation. The former one is the same as preludes radial Rashba components in untwisted or 30$^\circ$ twisted graphene/TMDC heterostructures\cite{Naimer2021:paper1,Zollner2023:PRB,Li2019:TwistTB1}. The latter one switches the sign of the in-plane spin components. Therefore, the combination of both is forbidding the emergence of a tangential (conventional) in-plane spin texture. If the  180$^\circ$-rotation symmetry is broken, both tangential and radial components are allowed. In the investigated materials, we find that this leads to mostly tangential spin textures at $K$ and mostly radial spin textures at $\Gamma$. The relevance of symmetries for the emergence of radial Rashba fields have been discussed in Refs.~\cite{MeraAcosta2021:3Dsymmetriesradial1,GosalbezMartinez2023:3Dsymmetriesradial2} and this symmetry in particular in Ref.~\cite{deSousa2025:radialRashba}.

One question that arises is the following: How can the occurrence of non-radial spin-orbit fields be described
by our model Hamiltonian? This can be done by adding a layer-dependent potential $H_{pot}=u\sigma_zs_0$. As we do not assume any external breaking of the symmetry (e.g. by an electric field) the source of it is the Moire potential induced by the twist\cite{Wu2017:PRL:twTMDC5}. However, this Moire potential is explicitly independent of the lateral shifting and also averages to zero. In supercells with 180$^\circ$-rotation symmetry the averaged spin textures we can observe with DFT therefore show no sign of this Moire potential and can hence be described with a model Hamiltonian without the additional $H_{pot}$. For supercells without this crucial symmetry, the asymmetry in the wave functions causes the Moire potential to not average out and result in a non-zero average effective potential difference.

\subsection{Twist angle dependency of the extracted parameters}
\begin{figure}
    \centering
    \includegraphics[width=0.79\linewidth]{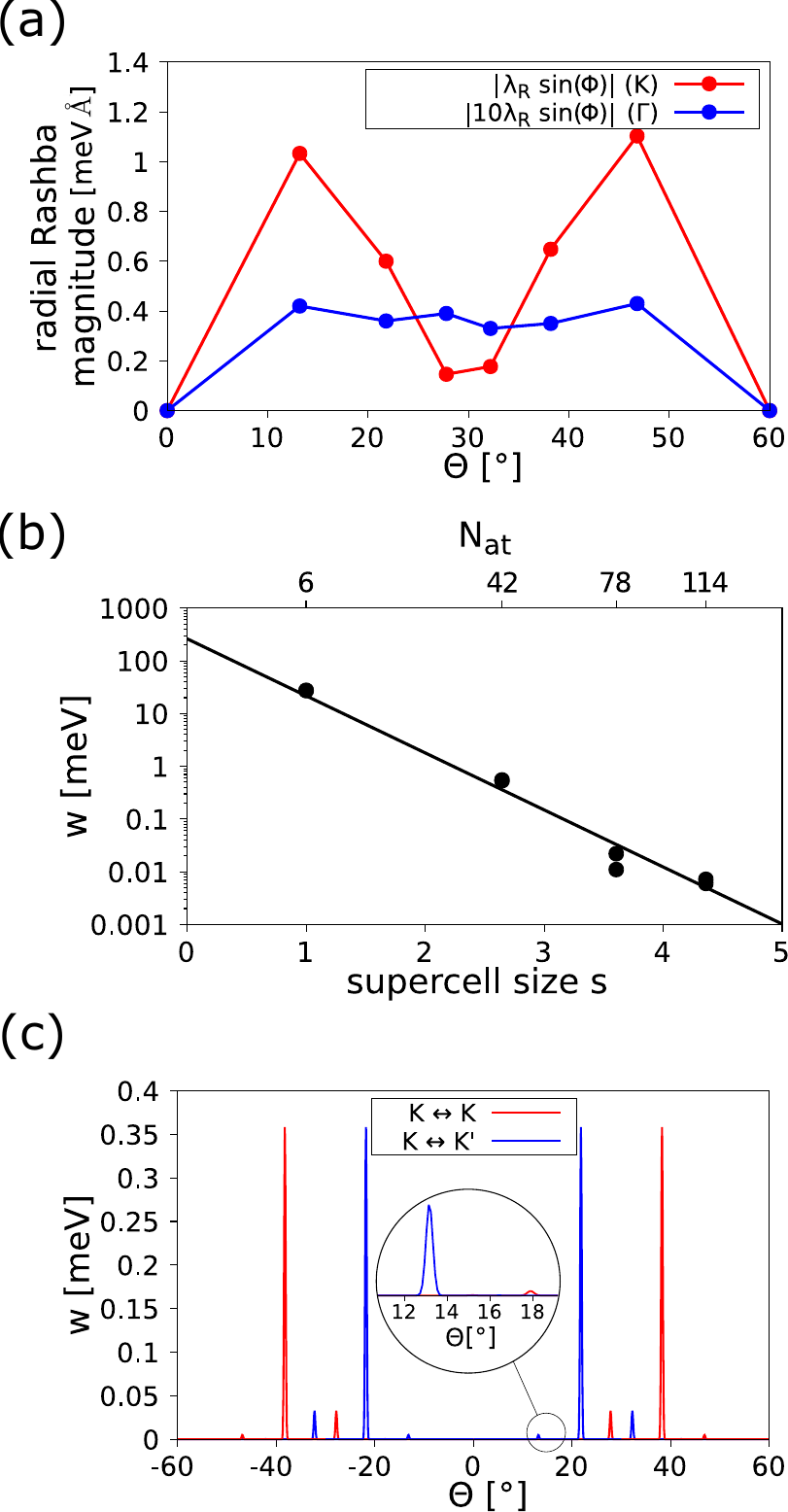}
    \caption{Model Hamiltonian parameters extracted from the $K$- and $\Gamma$-bands from the DFT calculations on twisted NbSe$_2$ bilayers. (a) Radial Rashba magnitude $|\lambda\sin(\Phi)|$ for $K$-bands (red) and $\Gamma$-bands (blue) as function of the twist angle $\Theta$. The lines are merely a guide to the eyes. The values of the $\Gamma$-bands are enhanced by a factor of 10 for better visibility. (b) Interlayer coupling $w$ as function of supercell size  (or $N_{at}$ of the supercell). The black line shows a fit, which is described in more detail in App.~\ref{App:relationtoconttw}. Only interlayer interactions drawn from the $K$-bands are shown, as the ones from the $\Gamma$-bands have a fixed value of $w=210$ meV (for a fixed interlayer distance). (c) Twist angle dependency of $w$ in the $K$-bands with the two backfolding scenarios color coded in blue ($K\leftrightarrow K'$) and red ($K\leftrightarrow K$). The inset shows a zoom to show another small peak at $\Theta\approx18^\circ$. The big peaks of interaction around the untwisted cases $\Theta=0^\circ$ and $\Theta=\pm60^\circ$ are omitted in order to focus on the twisted cases.}
    \label{Fig:extractedparams}
\end{figure}

For the NbSe$_2$ bilayers, we calculated a total of six different twist angles, in pairs of two, which are always partner angles. From their spin-orbit fields (see Fig.~\ref{Fig:SOF} (b)) we extract parameters using the model Hamiltonian from Eq.~\ref{Eq:Ham}. The results are listed in  Tab.~\ref{Tab:results} and shown in Fig.~\ref{Fig:extractedparams}. The relevant parameters to extract are the interlayer coupling $w$, the Rashba phase angle $\Phi$ and the magnitude of the total (hidden) Rashba $\lambda_R$. However, $\Phi$ and $\lambda_R$ cannot easily be extracted separately from each other, but only together in the form of $\lambda_R\sin(\Phi)$. Henceforth, we will call this the magnitude of the radial Rashba, which is not to be confused with the magnitude of the in-plane spin texture, although the two mostly overlap. 
Moreover, we cannot reliably extract the sign of $\lambda_R\sin(\Phi)$, due to deviations of the DFT data from the model predictions. {To be more precise, the model would predict the same ordering of the in-plane spins for band pairs VB1/2 and VB3/4 for the case $K\leftrightarrow K'$, $w>\sin(\Phi)\lambda_Rk$ (see Fig.~\ref{Fig:model} second column), i.e. the lower-energy band (VB2 and VB4) should show the same direction of in-plane spin. However, in our DFT calculations, this is not the case. Likely, the reason for this is the fact that the splittings are very small and prone to influence by nearby bands. Conversely, the data for large k-radii ($w<\sin(\Phi)\lambda_Rk$) agrees with the model for most cases and would hence allow an extraction of the sign. However, for the $\Theta=-32.2^\circ$ case, this data is also very noisy and does not represent a clear radial structure. Furthermore, for larger k-radii the model is naturally less applicable.}
Therefore, we refer simply to $|\lambda_R\sin(\Phi)|$.

In Fig.~\ref{Fig:extractedparams} (a) we show the twist angle dependence of the radial Rashba magnitude $|\lambda_R\sin(\Phi)|$. At $\Theta=0^\circ$ and $\Theta=60^\circ$ (untwisted cases) the radial Rashba vanishes by symmetry. Within this range $0^\circ<\Theta<60^\circ$ it should show a unique behavior that can be expanded to the whole range of 360$^\circ$ using the symmetry relations in Eq.~\ref{Eq:sym} and Eq.~\ref{Eq:sym2}. However, there appears to be an additional approximate symmetry around $\Theta=30^\circ$, connecting the partner angles, e.g. $\pm21.8^\circ$ and $\pm38.2^\circ$. {If one were to use the large k-radii in order to determine the sign of $\lambda_R\sin(\Phi)$, this would show an odd behavior of $\lambda_R\sin(\Phi)$ with respect to this symmetry axis.} The $\Gamma$-bands show a significantly smaller magnitude of the radial Rashba.

\begin{table}[htb]
   \caption{Model Hamiltonian parameters extracted from the $K$- and $\Gamma$-bands from the DFT calculations on twisted NbSe$_2$ bilayers. We list the twist angle $\Theta$, backfolding case (relevant for $K$-bands), interlayer coupling $w$ and the magnitude of the radial Rashba $|\lambda_R\sin(\Phi)|$.}
    \label{Tab:results} 
    \begin{ruledtabular}
    \begin{tabular}{c|c|cc|cc}
\multicolumn{2}{c}{NbSe$_2$/NbSe$_2$}&\multicolumn{2}{c}{$K$ bands}&\multicolumn{2}{c}{$\Gamma$ bands}\\
\hline
$\Theta[\degree]$&case&$w$&$|\lambda_{R}\sin(\Phi)|$&$w$&$|\lambda_{R}\sin(\Phi)|$ \\
$[\degree]$&& [meV]& [meV\AA]& [meV]& [meV\AA]\\

\hline
0& $K\leftrightarrow K$& 27.457 & 0.000 	& 243.250 & 0.000\\
21.8& $K\leftrightarrow K'$& 0.532 & 0.600 	& 209.972 & 0.036 \\

-38.2& $K\leftrightarrow K$& 0.551 & 0.648 	& 210.038 & 0.035 \\

-32.2& $K\leftrightarrow K'$& 0.011 & 0.177		& 210.148 & 0.033 \\

27.8& $K\leftrightarrow K$& 0.022& 0.146	& 210.086 & 0.039 \\

-46.8& $K\leftrightarrow K$& 0.007 & 1.103   &  210.102 & 0.043 \\

13.2& $K\leftrightarrow K'$& 0.006 & 1.033 & 210.119 & 0.042 \\
 
    \end{tabular}
    \end{ruledtabular}
    \end{table}

Now, let us discuss the interlayer coupling $w$ as extracted from the model Hamiltonian fit. The extracted $w$ for the $\Gamma$-bands all have nearly the same value of $w\approx210$~meV. 
The $w$ extracted from the $K$-bands can be found in Fig.~\ref{Fig:extractedparams} (b).  Here, we plot $w$ not against the twist angle $\Theta$, but rather against the size of the used supercell and find that with increasing size of the supercell the interaction between the $K$-bands decays exponentially. The explanation for this consists of two steps: First, the  condition for finding a small supercell in real space for a certain twist angle is the same as the condition for finding a point close to $\Gamma$ in extended $k$-space, where $K/K'$ of layer 1 and  $K/K'$ of layer 2 overlap. Second, the closer this overlapping point is to $\Gamma$, the stronger will be the interlayer coupling between these two points\cite{Koshino2015:TwistTBBasic}. The latter can also be seen as the reason for why the $w$ extracted from the $\Gamma$-bands is always a fixed large value (irrespective of the twist angle or supercell size), as here the interaction always occurs directly at $\Gamma$.
We stress that this is not a dependence on the system size per se, rather it is a dependence on the size of the smallest commensurate supercell that can be built at a certain twist angle. Hence -- considering a continuous change of the twist angle --  the interlayer coupling $w$ will peak at certain angles, where small commensurate supercells (like the ones we use in DFT) are possible, as shown in Fig.\ref{Fig:extractedparams} (c).
This argumentation is layed out in more detail in App.~\ref{App:relationtoconttw}.

\section{WT\lowercase{e}$_2$}
\label{Sec:WTe}
\begin{figure}
    \centering
    \includegraphics[width=0.99\linewidth]{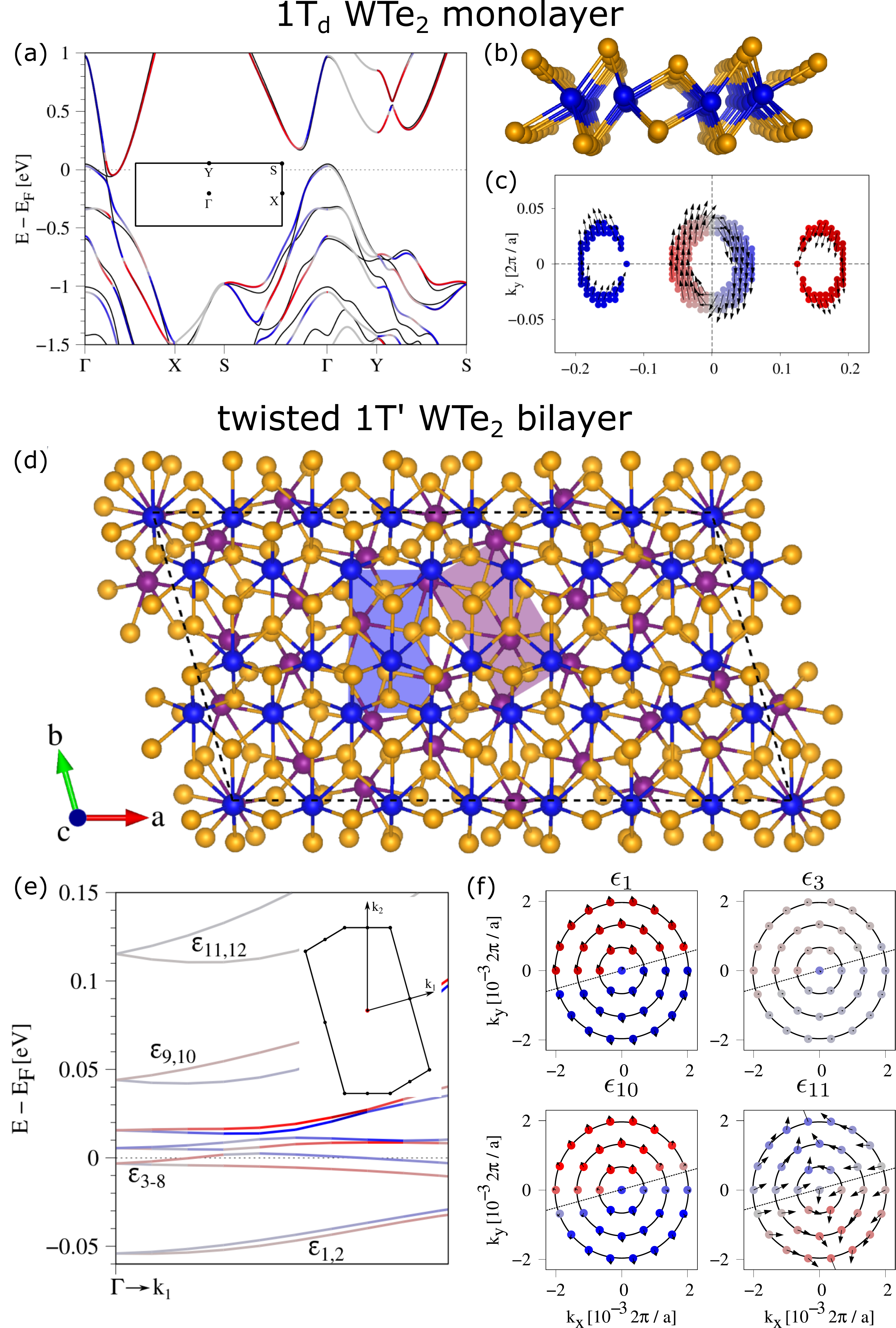}
    \caption{Real space structures, band structures and spin-orbit fields of 1T$_d$-WTe$_2$ monolayer (a)-(c) and twisted 1T'-WTe$_2$ bilayer (d)-(f). (a) Band structure (with color-coded spin-$z$ expectation values) of 1T$_d$-WTe$_2$ monolayer with FBZ as inset. (b) Side view of 1T$_d$-WTe$_2$ monolayer. (c) Spin-orbit fields of 1T$_d$-WTe$_2$ monolayer around $\Gamma$-point hole and nearby electron pockets. (d) Top view of the twisted 1T'-WTe$_2$ bilayer with monolayer unit cells of the two layers shown as purple and blue rectangles. W-atoms of the two layers are also shown in different colors (also blue and purple) for better distinction. (e) Band structure (with color-coded spin-$z$ expectation values) around $\Gamma$ of  bands ($\varepsilon_1$ to $\varepsilon_{12}$) close to the Fermi level of the twisted 1T'-WTe$_2$ bilayer with FBZ zone as inset. (f) Spin-orbit fields of some of the bands shown in (e). Again, spin-$z$ expectation values are color coded. The mirror symmetry axis is shown as dotted line. {In contrast to Fig.~\ref{Fig:SOF}, here spin-orbit fields are shown using only one band per plot and also with in-plane magnitude of the spins encoded in the length of the arrows.}}
    \label{Fig:WTe}
\end{figure}

In addition to WSe$_2$ and NbSe$_2$, we investigate the properties of bilayers of 1T'-WTe$_2$\cite{xu2018electrically, tang2017quantum, zhou2016pressure}.  This case requires a separate discussion, as it is different from the previously discussed ones in two major ways. Firstly, its monolayer's unit cell is not hexagonal, but orthorhombic. This complicates the formation of a commensurate twisted supercell, which requires some uniaxial strain. Contrary to its hexagonal counterparts, it does not have a $K$-point with parabolic bands. Instead, its main features close to the Fermi level and hence relevant for our discussion are the states directly around $\Gamma$ and the Fermi pockets located further along the $k_x$ direction (see Fig.~\ref{Fig:WTe} (a)).
Furthermore, WTe$_2$  is generally less symmetric than the other discussed TMDCs, especially lacking the typical $C_3$ symmetry.

As the monolayer of 1T'-WTe$_2$ is inversion-symmetric\cite{xu2018electrically}, we instead use the 1T$_d$ phase to illustrate the monolayer properties (see Fig.~\ref{Fig:WTe} (a)-(c)).
Monolayer 1T$_d$-WTe$_2$ has a vertical mirror plane and will hence not support a radial in-plane spin texture. Our calculations show that its spin texture close to the Fermi level is roughly tangential around $\Gamma$ and approximately uniform in the nearby pockets (see Fig.~\ref{Fig:WTe} (c)).

Moving along to the twisted bilayer, we used one supercell with a twist angle of $\Theta=31.14^\circ$ and a small strain of $\epsilon<0.1\%$ (see Fig.~\ref{Fig:WTe} (d)). Our DFT calculations find multiple bands close to the Fermi level at $\Gamma$ {(Fig.~\ref{Fig:WTe}~(e))}. When analyzing the spin-orbit fields of these bands, they are all symmetrical around the $\mathbf{k_1}$-axis, which is defined by the twist angle between the layers (dotted lines in Fig.~\ref{Fig:WTe} (f)). This is the consequence of the in-plane 180$^\circ$ rotation symmetry. As for the in-plane spin textures, a variety of different forms emerge. Some are approximately uniform, others are a mixture of uniform and radial spin textures, and some are a Dresselhaus-like (radial tangential) form. The emergence of a radial tangential spin texture due to a lack of C$_3$ symmetry has been predicted in Ref.~\cite{deSousa2025:radialRashba}. Analytical description of the 
spin textures is beyond the simple model that we used for 2H structures.

\section{Summary}
We performed first principles calculations on different twisted TMDC homobilayers, covering different materials (NbSe$_2$, WSe$_2$ and WTe$_2$), twist angles, lateral shifts and interlayer distances. Our calculations reveal that purely radial in-plane spin textures appear both around the $K$- and $\Gamma$-bands. They can be explained using a model Hamiltonian designed to describe commensurate supercells at large twist angles. We find the twist angle dependency of the relevant model parameters interlayer coupling $w$ and radial Rashba magnitude $|\lambda_R\sin(\Phi)|$, by fitting the model to the DFT data for different twist angles. Furthermore, we identify an in-plane 180$^\circ$ rotation symmetry as the crucial symmetry for upholding the purely radial spin textures. The investigations on non C$_3$-symmetric 1T' WTe$_2$ bilayers reveal novel spin-orbit fields, which are characterized by a vertical mirror symmetry, which is the sole symmetry of the system.

\acknowledgments All authors acknowledge support by the FLAG ERA JTC 2021 project 2DSOTECH, the European Union Graphene Flagship project 2DSPIN-TECH (grant agreement No. 101135853) and SFB 1277 (Project-ID 314695032). {The authors thank Martin Gmitra for fruitful discussion.}

\appendix

\section{Relation to continuous twist angle}
\label{App:relationtoconttw}
\begin{figure*}
    \centering
    \includegraphics[width=0.99\linewidth]{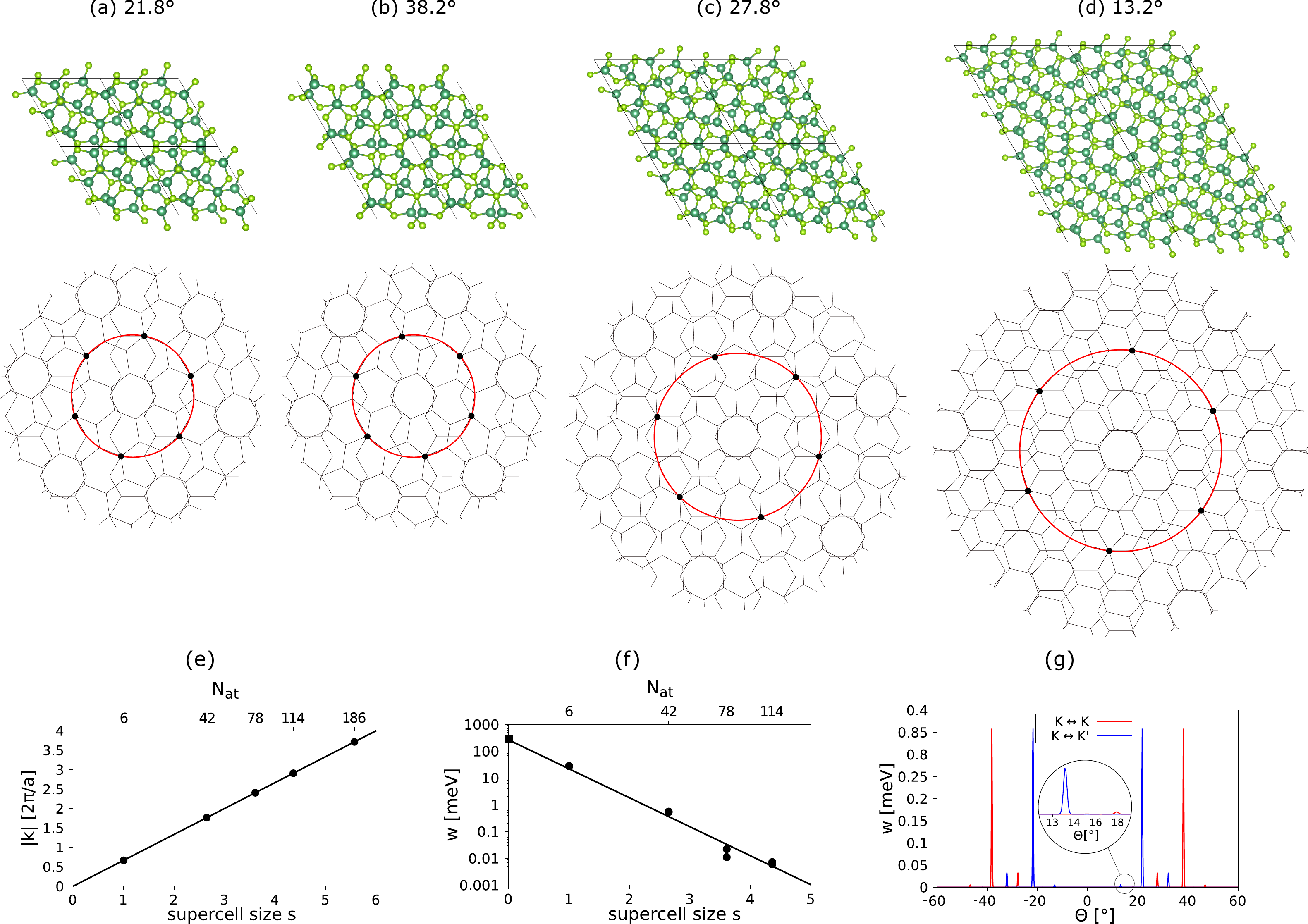}
    \caption{Establishing a realistic twist angle dependence of the interlayer coupling $w$ for a continuous twist angle. (a)-(d) For four different twist angles, we show the supercell and the corresponding overlapping extended Brillouin zones of the two layers. The red ring shows the $k$ radius $|k|$ at which the overlap of $K/K'$-points (black dots)) of the two layers occur. (e) Distance $|k|$ of the overlapping $K$-points to $\Gamma$ for differently sized supercells following a linear trend.
    (f) Extracted interlayer coupling $w$ as function of the supercell size (or $N_{at}$). Similar to Fig.~\ref{Fig:extractedparams} (b), but with additional data point at imaginary supercell size 0 with $w=210$meV, coming from the extracted $w$ at the $\Gamma$-bands. The fit is modeled with the function given in Eq.~\ref{Eq:fit}. (g) Twist angle dependence of $w$. Each peak stems from a possible small commensurate supercell; the smaller the supercell, the larger the peak. We additionally differentiate between peaks coming from $K\leftrightarrow K$ (red) and $K\leftrightarrow K'$ (blue) backfolding. The inset shows a zoom to show another small peak at $\Theta\approx18^\circ$. The big peaks of interaction around the untwisted cases $\Theta=0^\circ$ and $\Theta=\pm60^\circ$ are omitted in order to focus on the twisted cases.}
    \label{Fig:continuoustwist}
\end{figure*}

In our DFT calculations, we are restricted to using commensurate homobilayer supercells. The construction of such commensurate supercells is only possible for discrete values of the twist angle $\Theta$ (see Fig.~\ref{Fig:strucsbackfolding}(e)). We have demonstrated that the results of these calculations (in particular the radial in-plane spin textures) can be fairly described by the model Hamiltonian we introduce in Sec.~\ref{Sec:modelHam}. 
Hence, the question arises, if a radial Rashba spin texture will also arise, when we vary the twist angle continuously instead of choosing certain special discrete twist angles. In principle, for every twist angle (except $\Theta=30^\circ$), one can find a commensurate supercell representing the twisted homobilayer. However, these supercells can be very large and hence might not fulfill the necessary conditions for the occurrence of radial Rashba. 
The two necessary ingredients for the radial Rashba are a finite radial Rashba magnitude $\lambda_R\sin(\Phi)$ and a finite interlayer coupling $w$ between the parabolic bands. While the former one seems to be ubiquitous for all twisted structures, the latter one might not be present throughout a continuous twist angle range.
For the $\Gamma$-bands we find a stable $w\approx210$~meV, more than enough to sustain a radial Rashba, throughout all twist angles. Hence, radial Rashba spin textures are expected to arise near $\Gamma$ for all nonzero twist angles. For the $K$-bands, however, we observed a decaying trend of $w$ with supercell size. Let us analyze this unintuitive relation. The argumentation consists of two parts: 
\begin{enumerate}
    \item The relation between supercell size and the $k$-space distance $|k|$ between $\Gamma$ and the overlapping $K/K'$-points
    \item The relation between $|k|$ and interlayer coupling $w$
\end{enumerate}
Firstly, let us discuss the first relation:
Commensurate twisted supercells are generated by twisting two aligned untwisted layers around a common lattice point, until (coincidentally) two lattice points of the two layers (other than the ones at the twisting axis) overlap somewhere. The edge points of the new twisted supercell are set by the twisting axis and the overlapping lattice points, defining the size of the supercell. If we twist the extended $k$-spaces of the two layers around the $\Gamma$-point by the same twist angle, we will necessarily also find a repeating pattern with the same size as the one in real space. Additionally, an overlap between $K/K'$-points of the two layers will occur at a fixed ratio of this repeating pattern. This is illustrated in Fig.~\ref{Fig:continuoustwist} (a)-(d). 
The distance $|k|$ between this overlapping $K/K'$-points and $\Gamma$ therefore is linear to the supercell size (see Fig.~\ref{Fig:continuoustwist} (e)).
Concretely, in the case of TMDC homobilayers (with six atoms in the smallest untwisted supercell) this means:
\begin{align}
    |\mathbf{k}|=\frac{2\pi}{a}\frac{2}{3}s=\frac{2\pi}{a}\frac{2}{3}\sqrt{\frac{N_{at}}{6}},
\end{align}
with supercell size $s=\frac{a_s}{a}$ and number of atoms $N_{at}$ in the supercell.
It should be mentioned here that this relies on the fact that we always use the smallest possible supercell; in principle using a $2\times2$ supercell of a twisted supercell would always yield the same physics and same value for $w$, while technically having a larger supercell size. Therefore, when we refer to a supercell size it means the smallest supercell size possible at that specific twist angle.

Now, for the second relation we refer to  Ref.~\cite{Koshino2015:TwistTBBasic}. Since the transfer integral $T(\mathbf{r})$ between two atoms generally decays with their distance, it can also be assumed that its Fourier transform $t(\mathbf{q})$ decays in $q=|\mathbf{q}|$. This, ultimately, is the reason why the Umklapp process involving states far away from $\Gamma$ are very weak and therefore the interlayer coupling $w$ between these $K/K'$-points is very small.
We estimate the interlayer coupling to scale exponentially with $q=|\mathbf{k}|$ and therefore have
\begin{align}
\label{Eq:fit}
    w=a\exp(-b|\mathbf{k}|)=a\exp\Big(-b\frac{2\pi}{a}\frac{2}{{3}}s\Big),
\end{align}
with fitting parameters $a,b\in\mathbb{R}$.
In Fig.~\ref{Fig:continuoustwist} (f) this function is fitted using the data points from our calculations. The data point at $N_{at}=|\mathbf{k}|=0$ corresponds to the bands at $\Gamma$, where interaction takes place at $|\mathbf{k}|=0$ and where we consequently find large values of $w$, rather then a real supercell.

Combining Fig.~\ref{Fig:continuoustwist} (f) and Fig.~\ref{Fig:strucsbackfolding}(e), we can present an estimate of the twist angle dependent interlayer coupling between the $K/K'$-bands in Fig.~\ref{Fig:continuoustwist} (g). The peaks correspond to the possible commensurate supercells and decrease with increasing size of the supercell. We assume the peaks for each commensurate supercell to have a certain width. In reality, the physics around the commensurate case could include a lattice relaxation (favoring commensurate lattices) or flat bands occurring as a "higher order" magic angle effect. Also, the possible deviation from the commensurate case will be dependent on the Fermi level (so the distance to the $K$-point) at which we measure the spins.
This plot clearly shows that the supercells we chose for the DFT calculations (out of convenience of smaller computational cost) are automatically the ones amongst all possible twisted cells with the strongest interlayer coupling.
Now another question arises: How small can the peak be, before the radial in-plane spin texture breaks down?
According to our model Hamiltonian, the radial Rashba  appears at arbitrarily small interaction strength. In reality, however, there will be a limit at which interactions with other nearby bands and small perturbations will destroy the radial Rashba. Ultimately, it is likely that radial Rashba spin textures can also be found at other twist angles than the ones we investigate. However, the magnitude and stability of the effect is by far the strongest for the commensurate cases shown in our paper. 

\section{Interlayer distance study \& atomic relaxation}
\begin{figure*}
    \centering
    \includegraphics[width=0.99\linewidth]{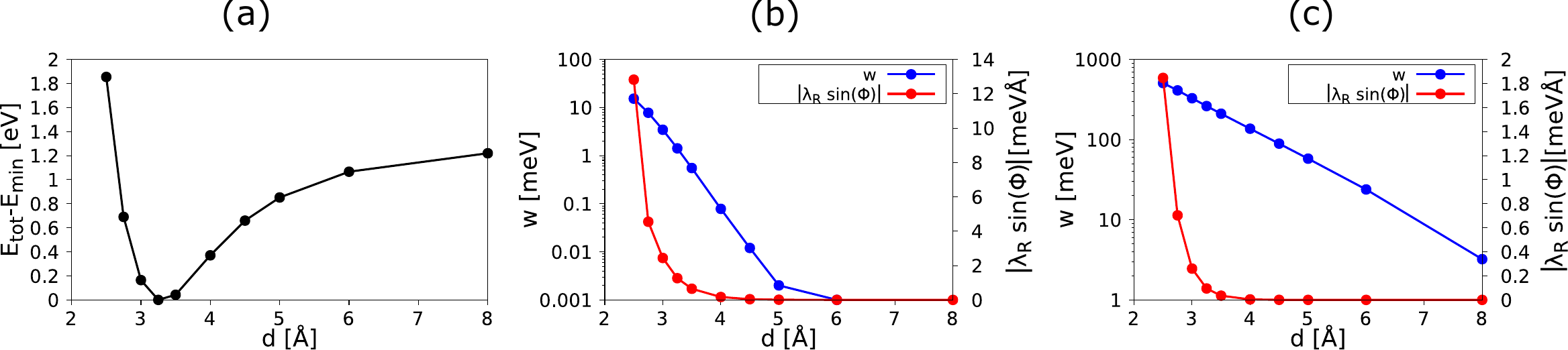}
    \caption{Interlayer distance study shows the dependence of several parameters on the interlayer distance $d$ between the layers. (a) Total energy $E_{tot}$ with respect to the smallest total energy found $E_{min}$. (b) Extracted parameters interlayer coupling $w$ and $|\lambda_R\sin(\Phi)|$ for the $K$-bands. (c) same as (b), but considering the $\Gamma$-bands. }
    \label{Fig:interlayerdist}
\end{figure*}

In order to further the understanding of the twisted homobilayers, we additionally performed an interlayer distance study for NbSe$_2$ bilayers with twist angle $\Theta=-38.2^\circ$. The interlayer distance is varied from $d=2.5$\AA ~to $d=8$\AA. We show the extracted parameters in Tab.~\ref{Tab:interlayer} and plot them in Fig.~\ref{Fig:interlayerdist}. We find that the interlayer coupling $w$  decays exponentially with $d$ for both the $K$- and the $\Gamma$-bands. Furthermore, the magnitude of the radial Rashba  $|\lambda_R\sin(\Phi)|$ decays even faster. The total energies of the systems seem to indicate that the equilibrium interlayer distance is at about $d=3.25$\AA.

This is congruent with the relaxation calculations we performed. Contrary to the other calculations throughout the main paper, we performed one calculation with prior atomic relaxation. The extracted parameters align rather well with the unrelaxed case with a comparable interlayer distance. The radial form of  the Rashba spin texture remains untouched.

\label{App:interlayerdist}

\begin{table}[htb]
   \caption{Parameters of supercells with different interlayer distances $d$ extracted from the band structure calculations using the model Hamiltonian.}
    \label{Tab:interlayer} 
    \begin{ruledtabular}
    \begin{tabular}{c|c|c|cc|cc}
\multicolumn{3}{c}{NbSe$_2$/NbSe$_2$}&\multicolumn{2}{c}{$K$ bands}&\multicolumn{2}{c}{$\Gamma$ bands}\\
\hline
$\Theta$&case&d&$w$&$|\lambda_{R}\sin(\Phi)|$&$w$&$|\lambda_{R}\sin(\Phi)|$\\
$[\degree]$&&[\AA]&[meV]&[meV\AA]&[meV]&[meV\AA]\\
\hline
-38.2& $K\leftrightarrow K$&2.5& 15.266 & 12.819 							& 506.730 & 1.846 \\
-38.2& $K\leftrightarrow K$&2.75& 7.766& 4.548  							& 410.184 & 0.703\\
-38.2& $K\leftrightarrow K$&3& 3.441 & 2.439  						& 327.569 & 0.260\\
-38.2& $K\leftrightarrow K$&3.25& 1.415 & 1.263  						& 261.612 & 0.095\\
-38.2& $K\leftrightarrow K$&3.5& 0.551 & 0.648 	                         & 210.038 & 0.035 \\
-38.2& $K\leftrightarrow K$&4& 0.078 & 0.167						& 136.559 & 0.004\\
-38.2& $K\leftrightarrow K$&4.5& 0.012 & 0.041						& 88.967 & 0.0004\\
-38.2& $K\leftrightarrow K$&5& 0.002 & 0.010 					& 57.741 & 0.0\\
-38.2& $K\leftrightarrow K$&6& 0.001 & 0.001 				& 23.842 & 0.0\\
-38.2& $K\leftrightarrow K$&8& 0.0 & 0.0 							& 3.220 & 0.0
    \end{tabular}
    \end{ruledtabular}
    \end{table}

\section{Computational Details}
\label{App:compdet}
\subsection{NbSe$_2$}
The electronic structure calculations on the NbSe$_2$ homobilayers were performed
implementing density functional theory (DFT)~\citep{Hohenberg1964:PRB} 
	using {\tt Quantum ESPRESSO}~\citep{Giannozzi2009:JPCM}.
	Self-consistent calculations are carried out with a $k$ point sampling of $30\times30$ (for $|\Theta|=21.8^\circ,38.2^\circ$), $9\times9$ (for $|\Theta|=27.8^\circ,32.2^\circ$) or $3\times3$ (for $|\Theta|=46.8^\circ,13.2^\circ$).
	We use charge density cutoffs $E_\rho=350$~Ry and wave function kinetic cutoff $E_{\text{wfc}}=60$~Ry for the fully relativistic pseudopotential
	with the projector augmented wave method~\citep{Kresse1999:PRB} with the 
	Perdew-Burke-Ernzerhof exchange correlation functional~\citep{Perdew1996:PRL}. We used Grimme D-2
    Van der Waals corrections~\citep{Grimme2006:JCC,Grimme2010:JCP,Barone2009:JCC}. The used lattice constant is $a_{\text{NbSe}_2}=3.26$\AA~and the used interlayer distance is $d=3.5$\AA. We added at least 19\AA~of vacuum to avoid interaction between the periodic images and therefore establish a quasi-2D system.
    
\subsection{WSe$_2$}

The \textit{ab initio} calculations of twisted WSe$_\text{2}$ homobilayers for two complementary twist angles are performed using \texttt{Wien2k}\cite{Wien2k}.   We employ the Perdew-Burke-Ernzerhof\cite{Perdew1996:PRL} exchange-correlation functional with van der Waals interactions included via the D3 correction\cite{Grimme2010:JCP}. We used a k-grid of $15 \times 15 \times 1$, and convergence criteria of 10$^{-6}$ $e$ for the charge and 10$^{-6}$ Ry for the energy. The plane-wave cutoff multiplied by the smallest atomic radii is set to 8. Spin–orbit coupling was included fully relativistically for core electrons, while valence electrons were treated within a second-variational procedure\cite{Singh2006:book} with the scalar-relativistic wave functions calculated in an energy window up to 5 Ry. The lattice parameter for the monolayer WSe$_\text{2}$ is $3.282 \; \textrm{\AA}$ and thickness is $3.34 \; \textrm{\AA}$\cite{FariaJunior2022NJP}. The interlayer distance is $3.4 \; \textrm{\AA}$\cite{Lin2021:Nat:TMDCexcitons}. The vacuum is $20 \; \textrm{\AA}$. To construct the twisted structures, we employed the {\tt Atomic Simulation Environment (ASE)}~\cite{ASE}, starting from two WSe$_2$ monolayers aligned at either $0^\circ$ or $60^\circ$ and subsequently applying the twisting procedure described in Ref.~\cite{Uchida2014:PRB}.

\subsection{1T'-WTe$_2$}
 The twisted WTe$_2$ homobilayer structure is set-up with the {\tt atomic simulation environment (ASE)} \cite{ASE} and the {\tt CellMatch} code \cite{Lazic2015:CPC}, implementing the coincidence lattice method \cite{Koda2016:JPCC,Carr2020:NRM}. 
 The lattice constants of pristine 1T'-WTe$_2$ are $a=3.48$~${\textrm{\AA}}$ and  $b=6.27$~${\textrm{\AA}}$. 
 Within the heterostructure, the top (bottom) 1T'-WTe$_2$ layer has lattice parameters of $a=3.48$~${\textrm{\AA}}$ ($a=3.479$~${\textrm{\AA}}$) and $b=6.27$~${\textrm{\AA}}$ ($b=6.271$~${\textrm{\AA}}$). 
 The involved strains are below 0.1\% and the twist angle between the layers is about 31.14°. 
 In order to simulate quasi-2D systems, we add a vacuum of about $20$~\AA~to avoid interactions between periodic images in our bilayer geometry. 
 The twisted 1T'-WTe$_2$ homobilayer  supercell has lattice vectors $|a| = 24.36$~\AA, $|b| = 13.0139$~\AA, and $|c| = 35.515$~\AA~ and an angle $\alpha = 105.51$°. The supercell contains 168 atoms, see Fig.~\ref{Fig:WTe}(d). 
 The electronic structure calculations and structural relaxations of the heterostructure are performed by DFT~\cite{Hohenberg1964:PRB} 
with {\tt Quantum ESPRESSO}~\cite{Giannozzi2009:JPCM}. Self-consistent calculations are carried out with a $k$-point sampling of $2\times 2\times 1$. 
We use an energy cutoff for charge density of $560$~Ry and the kinetic energy cutoff for wavefunctions is $70$~Ry for the fully relativistic pseudopotentials
with the projector augmented wave method~\cite{Kresse1999:PRB} with the 
Perdew-Burke-Ernzerhof exchange correlation functional~\cite{Perdew1996:PRL}.
For the self-consistent calculation, we employ a threshold of $1\times10^{-7}$~Ry and Fermi-Dirac smearing of $5\times10^{-4}$~Ry.
For the relaxation of the heterostructures, we add DFT-D2 vdW corrections~\cite{Grimme2006:JCC,Grimme2010:JCP,Barone2009:JCC} and use 
quasi-Newton algorithm based on trust radius procedure. 
To get proper interlayer distances and to capture possible moir\'{e} reconstructions, we allow all atoms to move freely within the heterostructure geometry during relaxation. Relaxation is performed until every component of each force is reduced below $1\times10^{-3}$~Ry/$a_0$, where $a_0$ is the Bohr radius.

\subsection{Fitting}

\begin{figure}
    \centering
    \includegraphics[width=0.99\linewidth]{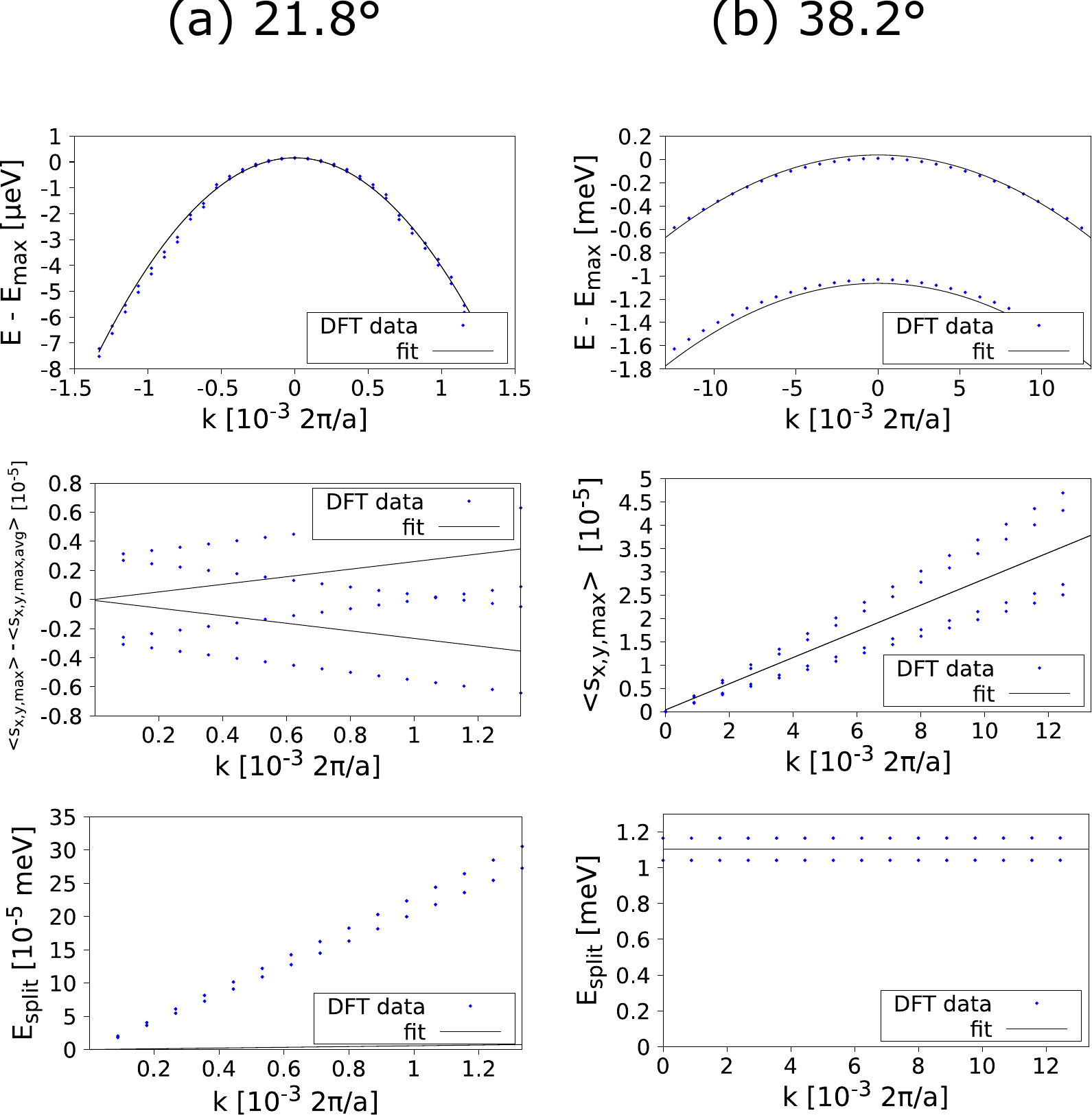}
    \caption{Comparison between model Hamiltonian fit and DFT. We compare energies $E$, magnitude of the radial in-plane spin textures (in form of the maximum in-plane spin expectation values $\langle s_{x,y,max}\rangle$) and energy splittings $E_{split}$ against the distance $k$ measured from the $K$-point for two cases (a) $|\Theta|=21.8^\circ$ and (b) $|\Theta|=38.2^\circ$. The two cases are representative of the two backfolding scenarios $K\leftrightarrow K'$ and $K\leftrightarrow K$, respectively. In the band structures the maximal energy of the 'valence bands' $E_{max}=E_F+632$meV is used as offset. For the spin expectation values, we use the offset $\langle s_{x,y,max,avg}\rangle\approx0.00346$ in the case of (a). The energy splitting for the 21.8$^\circ$ case is below the intended numerical accuracy 
     (meV), so the deviation between the model and DFT are not relevant.}
   
    \label{Fig:fitting}
\end{figure}

In the main paper we use a model Hamiltonian (see Eq.~\ref{Eq:Ham}) to extract relevant parameters from the DFT data by fitting. In Fig.~\ref{Fig:fitting} we compare the fit and DFT data for two examples ($K$-bands of NbSe$_2$ 21.8$^\circ$ and -38.2$^\circ$). To this end, we plot energies, magnitude of the radial in-plane spin textures and energy splittings against the distance $k$ measured from the $K$-point. We can see that, while the properties are generally well-reproduced by the Hamiltonian, there are a few problems. Firstly, both the radial spin texture's magnitude as well as the energy splittings are slightly different for VB1/2 and VB3/4, respectively. Therefore, the parameters we extract present an average of them. Furthermore, for $K\leftrightarrow K'$ backfolding (e.g. $\Theta=21.8^\circ$ case), although the energy splittings are linear in $k$ (as predicted by the model), the slope of the splitting with increasing $k$ is not congruent with the model predictions. One could focus the fit only on energies rather than spin expectation values in order to reproduce the splittings. However, the resulting parameters would not be realistic. Hence, in order to reproduce the relevant physics (in-plane spins), we neglect the ($\mu$eV) splittings in our fitting procedure for these cases.

\bibliography{references}

\end{document}